%% file: paper.tex
\documentclass[12pt]{article}
\usepackage{epsf}

\usepackage[hyperindex]{hyperref}

\input rgmac.latex

\newcommand\MSbar{\ensuremath{\overline{\rm MS}}}
\renewcommand\MeV{\ensuremath{{\rm MeV}}}
\renewcommand\GeV{\ensuremath{{\rm GeV}}}
\newcommand\CPT{{$\chi$}PT}
\newcommand\Dslash{{\setbox0=\hbox{$D$}\hbox to 0pt{\hbox to \wd0{\hfil/\hfil}\hss}\box0}}
\newcommand\pslash{{\setbox0=\hbox{$p$}\hbox to 0pt{\hbox to \wd0{\hfil/\hfil}\hss}\box0}}

\newcommand\mbar{\hbox{$\overline{m}$}}

\def\gsim{\mathrel{\raise2pt\hbox to 8pt{\raise -5pt\hbox{$\sim$}\hss{$>$}}}}
\def\rsim{\mathrel{\raise2pt\hbox to 8pt{\raise -5pt\hbox{$\sim$}\hss{$>$}}}}
\def\lsim{\mathrel{\raise2pt\hbox to 8pt{\raise -5pt\hbox{$\sim$}\hss{$<$}}}}
\def\vev#1{\langle #1 \rangle}

\begin{document}

\begin{titlepage}
 \null
 \begin{center}
 \makebox[\textwidth][r]{LAUR-98-271}
 \par\vspace{0.25in} 
  {\Large
       Quark Masses, B-parameters, and CP violation parameters $\epsilon$ and 
       $\epsilon'/\epsilon$
  }
  \par
 \vskip 2.0em
 
 {\large 
  \begin{tabular}[t]{c}
         Rajan Gupta \footnotemark\\[0.5em]
        \em Group T-8, Mail Stop B-285, Los Alamos National Laboratory\\
        \em Los Alamos, NM 87545, U.~S.~A\\[1.5em]
        \em Email:    rajan@qcd.lanl.gov\\[1.5em]
  \end{tabular}}
 \par \vskip 2.0em
 {\large\bf Abstract}
\end{center}

\quotation After a brief introduction to lattice QCD, I
summarize the results for the light quark masses and the bag
parameters $B_K$, $B_6^{1/2}$, and $B_8^{3/2}$.  The implications of
these results for the standard model estimates of CP violation
parameters $\epsilon$ and $\epsilon'/\epsilon$ are also discussed.

{
 \footnotetext {Based on invited talks given at the 
XVI AUTUMN  SCHOOL AND WORKSHOP ON FERMION MASSES, MIXING AND CP VIOLATION, 
Instituto Superior T\'ecnico, Lisboa, Portugal, 6-15 October 1997; and 
ORBIS SCIENTIAE 1997-II, PHYSICS OF MASS, Miami, Florida, Dec 12-15, 1997.}
}
\vfill
\mbox{20 JAN, 1998}
\end{titlepage}

\setlength{\textfloatsep}{12pt plus 2pt minus 2pt}

\makeatletter 

\setlength{\leftmargini}{\parindent}
\def\@listi{\leftmargin\leftmargini
            \topsep 0\p@ plus2\p@ minus2\p@\parsep 0\p@ plus\p@ minus\p@
            \itemsep \parsep}
\long\def\@maketablecaption#1#2{#1. #2\par}

\advance \parskip by 0pt plus 1pt minus 0pt

\makeatother

\section{Introduction}

The least well quantified parameters of the Standard Model (SM) are
the masses of light quarks and the $\rho$ and $\eta$ parameters in the
Wolfenstein representation of the CKM mixing matrix. A non-zero value
of $\eta$ signals CP violation.  The important question is whether the
CKM ansatz explains all observed CP violation. This can be addressed
by comparing the SM estimates of the two CP violating parameters
$\epsilon$ and $\epsilon'/\epsilon$ against experimental
measurements. The focus of this talk is to evaluate the dependence of
these parameters on the light quark masses and on the bag parameters
$B_K$, $B_6^{1/2}$, and $B_8^{3/2}$. I will therefore provide a status
report on the estimates of these quantities from lattice QCD (LQCD).

Since this is the only lecture presenting results obtained
using LQCD at this school/workshop, I have been asked to give some
introduction to the subject. The only way I can cover my charter,
introduce LQCD, summarize the results, and make contact with
phenomenology is to skip details.  I shall try to overcome this
shortcoming by giving adequate pointers to relevant literature.

\section{Lattice QCD}

LQCD calculations are a non-perturbative implementation of field theory
using the Feynman path integral approach. The calculations proceed
exactly as if the field theory was being solved analytically had we
the ability to do the calculations.  The starting point is the
partition function in Euclidean space-time
\begin{equation}
Z \ = \
\int {\cal D}A_{\mu} \ {\cal D}\psi \ {\cal D}\bar \psi \ e^{-S}
\end{equation}
where $S$ is the QCD action 
\begin{equation}
S \ = \ \int d^4 x \ \big( {1\over 4} F_{\mu\nu}F^{\mu\nu} -
                            \bar \psi M \psi \big) \ .
\end{equation}
and $M$ is the Dirac operator. The fermions are represented by Grassmann 
variables $\psi$ and $\bar \psi$. These can be integrated out exactly with 
the result 
\begin{equation}
Z \ = \
\int {\cal D}A_{\mu} \ {\rm det}M\ e^{\int d^4 x \ (-{1\over 4} F_{\mu\nu}F^{\mu\nu})} .
\end{equation}
The fermionic contribution is now contained in the highly non-local
term ${\rm det}M $, and the partition function is an integral over
only background gauge configurations.  
One can write the action, after integration over the fermions, as 
$S \ = {S_{gauge} + S_{quarks}} = 
       \ \int d^4 x \ \big( {1\over 4} F_{\mu\nu}F^{\mu\nu} \big)  -
                            \sum_i {\rm Ln}( {\rm Det} M_i)  $
where the sum is over the quark flavors 
distinguished by the value of the bare quark mass. Results for physical
observables are obtained by calculating expectation values
\begin{equation}
\vev{{\cal O}} \ = \ {1 \over Z}
\int {\cal D} A_{\mu} \ {\cal O} \ e^{-S} \ .
\label{eq:expvalue}
\end{equation}
where ${\cal O}$ is any given combination of operators expressed in
terms of time-ordered products of gauge and quark fields. The quarks
fields in $\CO$ are, in practice, re-expressed in terms of quark
propagators using Wick's theorem for contracting fields. In this way
all dependence on quarks as dynamical fields is removed. The basic
building block for the fermionic quantities, the Feynman propagator,
is given by
\begin{equation}
S_F(y,j,b;x,i,a) \ = \ \big( M^{-1} \big)^{y,j,b}_{x,i,a} \, ,
\end{equation}
where $M^{-1}$ is the inverse of the Dirac operator calculated on a
given background field. A given element of this matrix $\big( M^{-1}
\big)^{y,j,b}_{x,i,a}$ is the amplitude for the propagation of a 
quark from site $x$ with spin-color $i,a$ to site-spin-color $y,j,b$. 

So far all of the above is standard field theory. The problem we face
in QCD is how to actually calculate these expectation values and how
to extract physical observables from these.  I will illustrate the
second part first by using as an example the mass and decay constant
of the pion.

Consider the 2-point correlation 
function, $\vev{0 | \sum_x {\cal O}_f(\vec x,
t) {\cal O}_i(\vec 0,0) |0 }$, where the operators ${\cal O}$ are chosen to be 
the fourth component of the axial current 
$ {\cal O}_f = {\cal O}_i = A_4 = \bar \psi \gamma_4 \gamma_5 \psi$ 
as these have 
a large coupling to the pion. The 2-point correlation
function then gives the amplitude for creating a state
with the quantum numbers of the pion out of the vacuum at space-time
point $0$ by the ``source'' operator $ {\cal O}_i $; the 
evolution of this state to the point $(\vec x,t)$ via the QCD
Hamiltonian; and finally the annihilation by the ``sink'' operator
${\cal O}_f $ at $(\vec x,t)$.  The
rules of quantum mechanics tell us that ${\cal O}_i$ will create a
state that is a linear combination of all possible eigenstates of the
Hamiltonian that have the same quantum numbers as the pion, $i.e.$ the pion, 
radial excitations of the pion, three pions in $J=0$ state, $\ldots$. The second 
rule is that 
on propagating for Euclidean time $t$, a given eigenstate with energy
$E$ picks up a weight $e^{-Et}$.  Thus, the 2-point function can be
written in terms of a sum over all possible intermediate states
\begin{equation}
\vev{0 | \sum_x {\cal O}_f(\vec x,t) {\cal O}_i(0) |0 } = \sum_n 
{\vev{0 | {\cal O}_f | n} \vev{n| {\cal O}_i | 0} \over 2 E_n} \ e^{-E_n t} \, .
\label{eq:2point}
\end{equation}
To study the properties of the pion at rest we need to isolate this state
from the sum over $n$. To do this, the first simplification is to use the
Fourier projection $\sum_{\vec x}$ as it restricts the sum over states to
just zero-momentum states, so $E_n \to M_n$. (Note that it is
sufficient to make the Fourier projection over either ${\cal O}_i$ or
${\cal O}_f$.)  The second step to isolate the pion, $i.e.$ project in
the energy, consists of a combination of two strategies. One, make a
clever choice of the operators ${\cal O}_i$ to limit the sum over
states to a single state (the ideal choice is to set ${\cal O}_i$
equal to the quantum mechanical wave-function of the pion), and two,
examine the large $t$ behavior of the 2-point function where only the
contribution of the lowest energy state that couples to ${\cal O}_i$
is significant due to the exponential damping. Then 
\begin{equation}
\vev{0 | \sum_x {\cal O}_f(x,t) {\cal O}_i(0) |0 } 
\ \ \eqinf\ \ \ \ 
{\vev{0 | {\cal O}_f | \pi} \vev{\pi| {\cal O}_i | 0} \over 2 M_\pi} \ e^{-M_\pi t} \, .
\label{eq:2point1}
\end{equation}
The right hand side is now a function of the two quantities we want
since $ \vev{0 | A_4 | \pi} = M_\pi f_\pi$. In this way, the mass and
the decay constant are extracted from the rate of exponential fall-off
in time and from the amplitude.

\begin{figure}[t] 
\vspace{9pt}
\hbox{\hskip15bp\epsfxsize=0.9\hsize \epsfbox {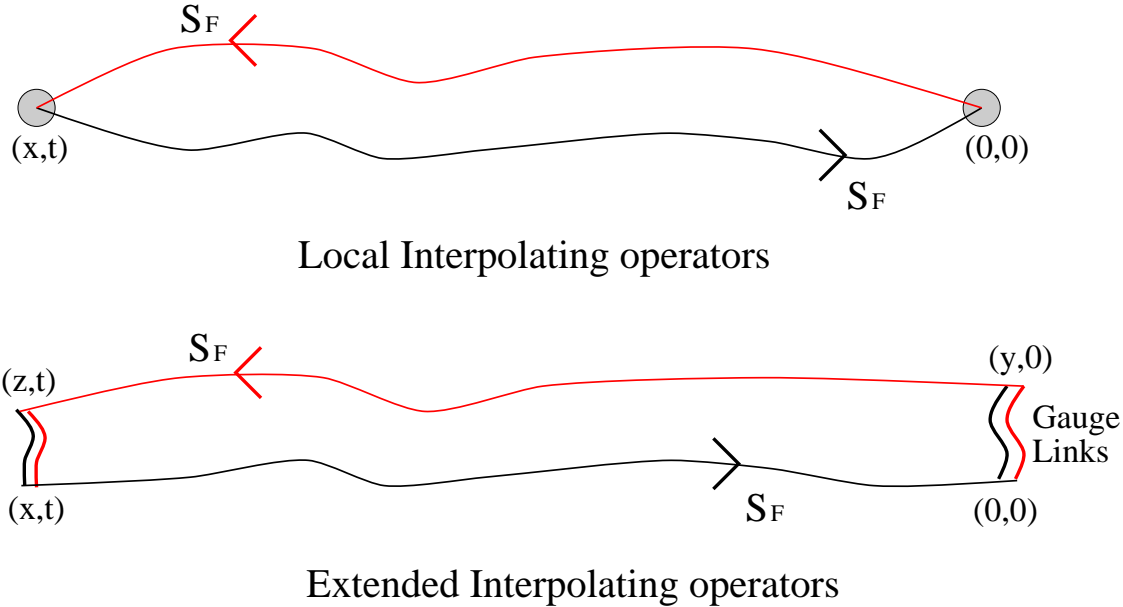}}
\caption{A schematic of the pion 2-point correlation function for local  
and non-local interpolating operators.}
\label{f:pionfig}
\end{figure}

Let me now illustrate how the left hand side is expressed in
terms of the two basic quantities we control in the path integral --
the gauge fields and the quark propagator.  Using Wick contractions,
the correlation function can be written in terms of a product of two
quark propagators $S_F$,
\begin{equation}
\vev{0 | \sum_x \bar \psi \gamma_4 \gamma_5 \psi(x,t) \bar \psi \gamma_4 \gamma_5 \psi(0,0) |0 } 
\ \equiv \ \vev{0 | \sum_x S_F(0; \vec x, t) \gamma_4\gamma_5 
                           S_F(\vec x, t; 0) \gamma_4 \gamma_5 | 0  } .
\label{eq:2point2}
\end{equation}
This correlation function is illustrated in Fig.~\ref{f:pionfig}. It
is important to note that one recovers the 2-point function
corresponding to the propagation of the physical pion only after the
functional integral over the gauge fields, as defined in
Eq.~\ref{eq:expvalue}, is done.  To illustrate this Wick contraction
procedure further, consider using gauge invariant non-local operators,
for example using $\CO = \bar \psi(x,t) \gamma_4 \gamma_5 (\CP
e^{\int^{y}_{x}dz igA_\mu(z)}) \psi(y,t)$ where $\CP$ stands for
path-ordered.  After Wick contraction the correlation function reads
\begin{equation}
\vev{0 | \sum_x S_F(0; \vec x, t)   \gamma_4\gamma_5 (\CP e^{\int^{x}_{z} igA_\mu})
                S_F ( \vec z, t; \vec y,0) \gamma_4\gamma_5 (\CP e^{\int^{y}_{0} igA_\mu}) | 0  } .
\label{eq:2point3}
\end{equation}
and involves both the gauge fields and quark propagators. This
correlation function would have the same long $t$ behavior as shown in
Eq.~\ref{eq:2point}, however, the amplitude will be different and
consequently its relation to $f_\pi$ will no longer be simple. The
idea of improving the projection of $\CO$ on to the pion is to construct
a suitable combination of such operators that approximates the pion
wave-function.

To implement such calculations of correlation functions requires the
following steps. A way of generating the background gauge
configurations and calculating the action $S$ associated with each;
calculating the Feynman propagator on such background fields;
constructing the desired correlation functions; doing the functional
integral over the gauge fields to get expectation values; making fits
to these expectation values, say as a function of $t$ as in
Eq.~\ref{eq:2point} to extract the mass and decay constant; and
finally including any renormalization factors needed to properly
define the physical quantity.  It turns out that at present the only
first principles approach that allows us to perform these steps is
LQCD.  Pedagogical expose to LQCD can be found in
\cite{Creutz83,Creutz92,Montvay,Rothe}, and I shall only give a very
brief description here.

Lattice QCD -- QCD defined on a finite space-time grid -- serves two
purposes.  One, the discrete space-time lattice serves as a
non-perturbative regularization scheme. At finite values of the
lattice spacing $a$, which provides the ultraviolet cutoff, there are
no infinities. Furthermore, renormalized physical quantities have a
finite well behaved limit as $a \to 0$.  Thus, in principle, one could
do all the standard perturbative calculations using lattice
regularization, however, these calculations are far more complicated
and have no advantage over those done in a continuum scheme.  The
pre-eminent utility of transcribing QCD on the lattice is that LQCD
can be simulated on the computer using methods analogous to those used
in Statistical Mechanics.  These simulations allow us to calculate
correlation functions of hadronic operators and matrix elements of any
operator between hadronic states in terms of the fundamental quark and
gluon degrees of freedom following the steps discussed above.

The only tunable input parameters in these simulations are the strong
coupling constant and the bare masses of the quarks.  Our belief is
that these parameters are prescribed by some yet more fundamental
underlying theory, however, within the context of the standard model
they have to be fixed in terms of an equal number of experimental
quantities. This is what is done in LQCD. Thereafter all predictions of
LQCD have to match experimental data if QCD is the correct theory of
strong interactions.

A summary of the main points in the calculations of expectation values
via simulations of LQCD are as follows.

\begin{itemize}

\item
The Yang-Mills action for gauge fields and the Dirac operator for
fermions has to be transcribed on to the discrete space-time lattice
in such a way as to preserve all the key properties of QCD --
confinement, asymptotic freedom, chiral symmetry, topology, and a
one-to-one relation between continuum and lattice fields. This step is
the most difficult, and even today we do not have a really
satisfactory lattice formulation that is chirally symmetric in the
$m_q=0$ limit and preserves the one-to-one relation between continuum
and lattice fields, $i.e.$ no doublers.  In fact, the Nielson-Ninomiya
theorem states that for a translationally invariant, local, hermitian
formulation of the lattice theory one cannot simultaneously have
chiral symmetry and no doublers \cite{NNtheorem}. One important
consequence of this theorem is that, in spite of tremendous effort, there is
no viable formulation of chiral fermions on the lattice. For a review
of the problems and attempts to solve them see
\cite{CGT95shamir,CGT97narayanan,CGT97testa}.  

A second problem is encountered when approximating derivatives in the
action by finite differences. As is well known this introduces
discretization errors proportional to the lattice spacing $a$. These
errors can be reduced by either using higher order difference schemes
with coefficients adjusted to take into account effects of
renormalization, or equivalently, by adding appropriate combinations of
irrelevant operators to the action that cancel the errors order by
order in $a$.  The various approaches to improving the fermion and
gauge actions are discussed in
\cite{IMP97lepage,IMP97alpha,IMP97hasenfratz}.  Here I simply list the
three most frequently used discretizations of the Dirac action --
Wilson \cite{Waction}, Sheikholeslami-Wohlert (clover)
\cite{SWaction}, and staggered \cite{STAGaction}, which have errors of
$O(a)$, $O(\alpha_s a)-O(a^2)$ depending on the value of the
coefficient of the clover term, and $O(a^2)$ respectively.  The
important point to note is that while there may not yet exist a perfect
action (no discretization errors) for finite $a$, improvement of the
action is very useful but not necessary.  Even the simplest
formulation, Wilson's original gauge and fermion action
\cite{Waction}, gives the correct results in the $a = 0$ limit. It is
sufficient to have the ability to reliably extrapolate to $a=0$ to
quantify and remove the discretization errors.

\item
The Euclidean action $S = \int d^4x (\ {1\over 4} F_{\mu\nu}F^{\mu\nu}
- {\rm Tr Ln } M) $ for QCD at zero chemical potential is real and
bounded from below.  Thus $e^{-S}$ in the path integral is analogous
to the Boltzmann factor in the partition function for statistical
mechanics systems, $i.e.$ it can be regarded as a
probability weight for generating configurations.  Since $S$ is an
extensive quantity the configurations that dominate the functional
integral are those that minimize the action. The ``importance
sampled'' configurations (configurations with probability of occurrence
given by the weight $e^{-S}$) can be generated by setting up a Markov
chain in exact analogy to say simulations of the Ising model. For a
discussion of the methods used to update the configurations see
\cite{Creutz83} or the lectures by Creutz and Sokal in
\cite{Creutz92}.

\item
The correlation functions are expressed as a product of quark
propagators and path ordered product of gauge fields using Wick
contractions. This part of the calculation is standard field theory.
The only twist is that the calculation is done in Euclidean
space-time.

\item
For a given background gauge configuration, the Feynman quark
propagator is a matrix labeled by three indices -- site, spin and
color. A given element of this matrix gives the amplitude for the
propagation of a quark with some spin, color, and space-time point to
another space-time point, spin, and color.  Operationally, it is
simply the inverse of the Dirac operator.  Once space-time is made
discrete and finite, the Dirac matrix is also finite and its inverse
can be calculated numerically.  The gauge fields live on links between
the sites with the identification $U_\mu(x, x + \hat \mu) =
e^{iagA_\mu(x+\hat \mu/2)}$, $i.e.$ the link at site $x$ in the $\mu$
direction is an SU(3) matrix $U_\mu(x,x + \hat \mu)$ denoting the
average gauge field between $x$ and $x + \hat \mu$ and labeled by the
point $x+\hat \mu/2$. Also $U_\mu(x,x - \hat \mu) \equiv
U^\dagger_\mu(x-\hat \mu,x)$.  The links and propagators can be
contracted to form gauge invariant correlation functions as discussed
above in the case of the pion.

\item
On the ``importance sampled'' configurations, the expectation values
reduce to simple averages of the correlation functions.  The problem
is that the set of background gauge configurations is infinite.  Thus,
while it is possible to calculate the correlation functions for
specified background gauge configurations, doing the functional
integral exactly is not feasible. It is, therefore, done numerically
using monte carlo methods.

\end{itemize}

The simplest way to understand the numerical aspects of LQCD
calculations is to gain familiarity with the numerical treatment of
any statistical mechanics system, for example the Ising model.  The
differences are: (i) the degrees of freedom in LQCD are much more
complicated -- SU(3) link matrices rather than Ising spins, and quark
propagators given by the inverse of the Dirac operator; (ii) The
action involves the highly nonlocal term ${\rm Ln\ Det\ } M$ which
makes the update of the gauge configurations very expensive; and (iii)
the correlation functions are not simple products of spin variables
like the specific heat or magnetic susceptibility, but complicated
functions of the link variables and quark propagators.

The subtleties arising due to the fact that LQCD is a renormalizable
field theory and not a classical statistical mechanics system come
into play in the behavior of the correlation functions as the lattice
spacing $a$ is varied, and in the quantum corrections that renormalize
the input parameters (quark and gluon masses and fields) and the
composite operators used in the study of correlation functions. At
first glance it might seem that one has introduced an additional
parameter in LQCD, the lattice spacing $a$, however, recall that the
coupling $\alpha_s$ and the cutoff $a$ are not independent quantities
but are related by the renormalization group
\begin{equation}
\Lambda_{QCD} = {1\over a} e^{-1/2\beta_0 g^2(a)}\ \big(\beta_0 g^2(a) \big)^{-\beta_1/2\beta_0^2} + \ldots \ ,
\label{eq:Lambdadef}
\end{equation}
where $\Lambda_{QCD}$ is the non-perturbative scale of QCD, and
$\beta_0 = (11 - 2n_f/3)/16\pi^2$ and $\beta_1 =
(102-38n_f/3)/(16\pi^2)^2$ are the first two, scheme independent, coefficients of 
the $\beta$-function.  In statistical mechanics systems, the
lattice spacing $a$ is a physical quantity -- the intermolecular
separation. In QFT it is simply the ultraviolet regulator that must
eventually be taken to zero keeping physical quantities, like the
renormalized coupling, spectrum, etc, fixed.

The reason that lattice results are not exact is because in numerical
simulations we have to make a number of approximations.  The size of
these is dictated by the computer power at hand. They are being
improved steadily with computer technology, better numerical
algorithms, and better theoretical understanding. To evaluate the
reliability of current lattice results, it is important to understand
the size of the various systematic errors and what is being done to
control them. I, therefore, consider it important to discuss these
next before moving on to results.

\section{Systematic Errors in Lattice Results}
\label{s:syserrors}

The various sources of errors in lattice calculations are as follows. 

{\bf Statistical errors:}
The monte carlo method for doing the functional integral employs
statistical sampling. The results, therefore, have statistical errors.
The current understanding, based on agreement of results from
ensembles generated using different algorithms and different initial
starting configuration in the Markov process, is that the functional
integral is dominated by a single global minimum. Also, configurations
with non-trivial topology are properly represented in an ensemble
generated using a Markov chain based on small changes to link
variables.  Another way of saying this is that the data indicate that
the energy landscape is simple.  As a result, the statistical accuracy
can be improved by simply generating more statistically independent
configurations with current update methods. 

{\bf Finite Size errors:}
Using a finite space-time volume with (anti-)periodic boundary
conditions introduces finite size effects. On sufficiently large
lattices these effects can be analyzed in terms of interactions of the
particle with its mirror images. L\"uscher has shown that in this
regime these effects vanish exponentially \cite{FSE88Luscher}.
Current estimates indicate that for $L \gsim 3$ fermi and $M_\pi L \ge
6$ the errors are $\lsim 1\%$, and decrease exponentially with increasing $L$. 

{\bf Discretization errors:} The discretization of the Euclidean
action on a finite discrete lattice with spacing $a$ leads, in
general, to errors proportional to $a$, $\alpha_s^n a$, $a^2$,
$\ldots$.  The precise form of the leading term depends on the choice
of the lattice action and operators \cite{IMPROVE}.  For example,
lattice artefacts in the fermion action modify the quark propagator
$M^{-1}$ at large $p$ from its continuum form. Numerical data show
that the coefficients of the leading term are large, consequently the
corrections for $1/a \approx 2 \GeV$ are significant in many
quantities, 10-30\% \cite{IMPROVEtest}.  The reliability of lattice
results, with respect to $O(a)$ errors, is being improved by a two
pronged strategy. First, for a given action extrapolations to the
continuum limit $a=0$ are performed by fitting data at a number of
values of $a$ using leading order corrections.  Second, these
extrapolations are being done for different types of actions (Wilson,
Clover, staggered) that have significantly different discretization
errors. We consider the consistency of the results in the $a=0$ limit
as a necessary check of the reliability of the results.

{\bf Extrapolations in Light Quark Masses:} The physical $u$ and $d$
quark masses are too light to simulate on current lattices. For $1/a =
2$ \GeV, realistic simulations require $L/a \gsim 90$ to avoid finite
volume effects, $i.e.$ keeping $\tilde M_\pi L \ge 6$ where $\tilde M_\pi$ is 
the lightest pseudoscalar meson mass on the lattice.  Current best
lattice sizes are $L/a=32$ for quenched and $L/a=24$ for unquenched. Thus,
to get results for quantities involving light quarks, one typically
extrapolates in $m_u = m_d$ from the range $ m_s/3 - 2 m_s$ using
simple polynomial fits based on chiral perturbation theory.  For
quenched simulations there are additional problems for $m_q \lsim
m_s/3$ as discussed below in the item on quenching errors.

{\bf Discretization of heavy quarks:} Simulations of heavy quarks ($c$
and $b$) have discretization errors of $O(ma)$ and $O(pa)$. This is
because quark masses measured in lattice units, $m_c a$ and $m_b a$,
are of order unity for $2 \GeV \le 1/a \le 5 \GeV$.  It turns out that
these discretization errors are large even for $m_c$. Extrapolations
of lattice data from lighter masses to $m_b $ using HQET have also not
been very reliable as the corrections are again large. The three most
promising approaches to control these errors are non-relativistic QCD,
$O(a)$ improved heavy Dirac, and HQET.  These are discussed in
\cite{NRQCD,heavyIMP,HQET}.  There will not be any discussion of 
heavy quark physics in this talk.

{\bf Matching between lattice and the continuum (renormalization
constants):} Experimental data are analyzed using some continuum
renormalization scheme like $\overline{MS}$, so results in the lattice
scheme have to be converted to this scheme.  The perturbative relation
between renormalized quantities in say $\overline{MS}$ and the lattice
scheme, are in almost all cases, known only to 1-loop.  Data show that
the $O(\alpha_s^2)$ corrections can be large, $\sim 10-50\%$ depending
on the quantity at hand, even after implementation of the improved
perturbation theory technique of
Lepage-Mackenzie~\cite{lepagemackenzie}. Recently, the technology to
calculate these factors non-perturbatively has been developed and is
now being exploited~\cite{nonpertZ}.  As a result, the reliance on
perturbation theory for these matching factors will be removed.

{\bf Operator mixing:} The lattice operators that arise in the
effective weak Hamiltonian can, in general, mix with operators of the
same, higher, and {\it lower} dimensions because at finite $a$ the
symmetries of the lattice and continuum theories are not the same.
Perturbative estimates of this mixing can have an even more serious
problem than the uncertainties discussed above in the matching
coefficients.  In cases where there is mixing with lower dimensional
operators, the mixing coefficients have to be known very accurately
otherwise the power divergences overwhelm the signal as $a \to 0$.  In
cases where there is mixing, due to the explicit chiral symmetry
breaking in Wilson like actions, with operators of the same dimension
but with different tensor structures, the chiral behavior may again
be completely overwhelmed by the artefacts. In both of these cases a
non-perturbative calculation of the mixing coefficients is essential.

{\bf Quenched approximation:} The inclusion of the fermionic
contribution, ${\rm Ln\ det}(M)$, in the Boltzmann factor $e^{-S}$ for
the generation of background gauge configurations increases the
computational cost by a factor of $10^3 - 10^4$.  The strategy,
therefore, has been to initially neglect this factor, and to bring all
other above mentioned sources of errors under quantitative control.
The justification is that the quenched theory retains a number of the
key features of QCD -- confinement, asymptotic freedom, and the
spontaneous breaking of chiral symmetry -- and is expected to be good
to within $10-20\%$ for a number of quantities. One serious drawback
is that the quenched theory is not unitary and \CPT\ analysis of it
shows the existence of unphysical singularities in the chiral
limit. For example, the chiral expansions of pseudoscalar masses and
decay constants in the quenched theory are modified in two ways. One,
the normal chiral coefficients are different in the quenched theory,
and second there are additional terms that are artefacts and are
singular in the limit $m_q = 0$ \cite{QCPTsrs,QCPTcbmg}.  These
artefacts are expected to start becoming significant for $ m_q \lsim
m_s/3$ \cite{QCPTtests,BK96srs}. Thus, in quenched simulations one of
the strategies for extrapolations in the light quark masses is to use
fits based on \CPT, keep only the normal coefficients, and restrict
the data to the range $ m_s/3 - 2 m_s$ where the artifacts are
expected to be small. In this talk I shall use this procedure to
``define'' the quenched results.

\medskip

The above mentioned systematic errors are under varying degrees of
control depending on the quantity at hand. Of the systematics effects
listed above, quenching errors are by far the least well quantified,
and are, to first approximation, unknown. Of the remaining sources the
two most serious are the discretization errors and the matching of
renormalized operators between the lattice and continuum theories.  An
example of the latter is the connection between the quark mass in
lattice scheme and in a perturbative scheme like $\overline{MS}$. We
shall discuss the status of control over these errors in more details
when discussing data.

\section{Light Quark Masses from \CPT}
\label{s:mfromcpt}

The masses of light quarks cannot directly be measured in experiments
as quarks are not asymptotic states.  One has to extract the masses
from the pattern of the observed hadron spectrum. Three approaches
have been used to estimate these -- chiral perturbation theory (\CPT),
QCD sum-rules, and lattice QCD.

\CPT\ relates pseudoscalar meson masses to \(m_u\), \(m_d\), and
\(m_s\). However, due to the presence of an overall unknown scale in
the chiral Lagrangian, \CPT\ can predict only two ratios amongst the
three light quark masses. The current estimates are 
\cite{gasserPR,rMq90leutwyler,rMq96leutwyler}

\medskip
\begin{center}
\begin{tabular}{|c|c|c|}
\hline
                          &  Lowest order & Next order \\
\hline
$ 2  m_s / (m_u + m_d) $  &  $24.2-25.9$  & $  24.4(1.5)  $  \\ 
$           m_u / m_d  $  &  $0.55  $     & $  0.553(43)  $ \ . \\
\hline
\end{tabular}
\label{tab:mqcpt}
\end{center}
\smallskip
These ratios have been calculated neglecting the Kaplan-Manohar
symmetry \cite{KaplanManohar}. The subtle point here is that the masses
$\mu_i$ extracted from low energy phenomenology are related to the
fundamental parameters $m_i$, defined at some high scale by the
underlying theory, as $\mu_i=m_i+\beta*\det M / m_i\Lambda_{\chi SB}$,
where the second, correction, term is an instanton induced additive
renormalization. For the $u$ quark, the magnitude of this term is
roughly equal to the \CPT\ estimates of $\mu_u$ for $\beta \approx 2$~\cite{Mq94Banks}.
Consequently, using \CPT, one cannot estimate the size of isospin
breaking from low energy phenomenology alone.  At next to leading order, only
one combination of ratios ($Q^2 = (m_s^2 - \mbar^2)/(m_d^2-m_u^2)$ as
defined in \cite{rMq96leutwyler}) can be determined unambiguously from
\CPT. Even if one ignores the Kaplan-Manohar subtlety (such an
approach has been discussed by Leutwyler under the assumption that the
higher order terms are small \cite{rMq96leutwyler}) one still needs
input from sum-rules or LQCD to get absolute values of quark masses.

\section{Light Quark Masses from LQCD}
\label{s:mfromlqcd}

\begin{figure}[htb] 
\vspace{9pt}
\hbox{\hskip15bp\epsfxsize=0.7\hsize \epsfbox {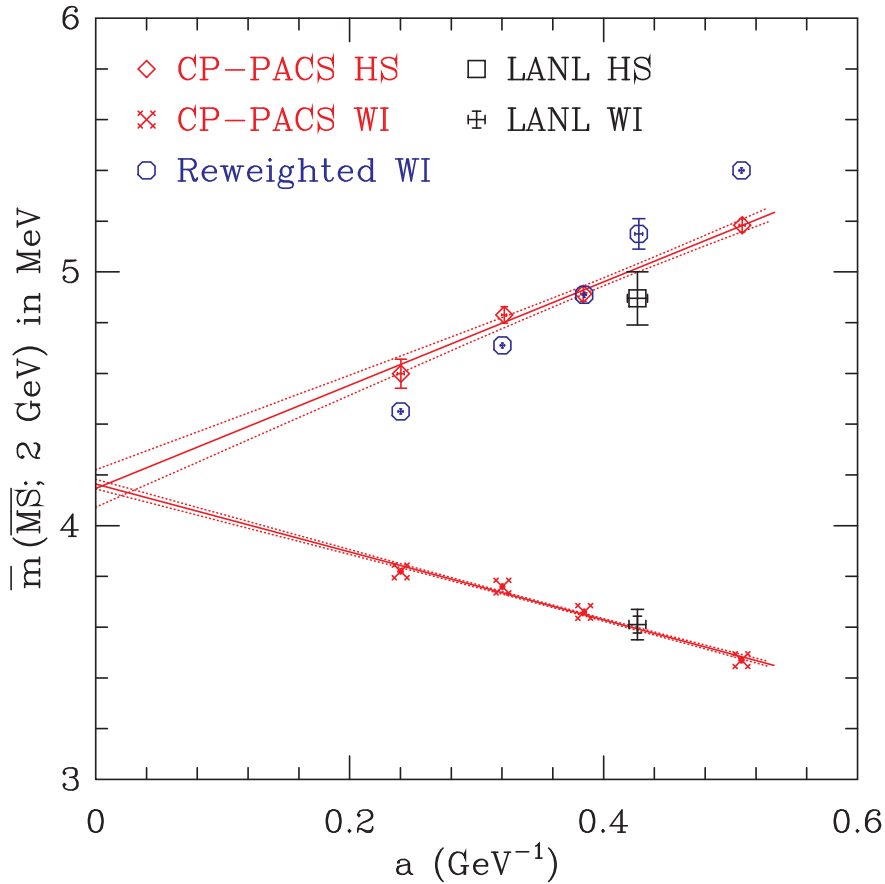}}
\caption{Linear extrapolation of $\mbar$ versus $a(M_\rho)$ for Wilson fermions 
using HS and WI methods. The WI data corrected by using non-perturbative estimates 
for the matching constants are also shown.}
\label{f:Wmbar}
\end{figure}

The most extensive and reliable results from LQCD have been obtained
in the quenched approximation. In the last year the statistical
quality of the quenched data has been improved dramatically especially by the
work of the two Japanese Collaborations CP-PACS and JLQCD (see \cite{Mq97rev} 
for a recent review).
Simultaneously, the lattice sizes have been pushed to $ \gsim 3$ fermi
for the lattice spacing in the range $0.5 - 0.25$ $\GeV^{-1}$. In
Fig.~\ref{f:Wmbar} we show the CP-PACS data obtained using Wilson
fermions.  To highlight the statistical improvement we show data at
$\beta=6.0$ from the next best calculation (with respect to both
statistics and lattice size) \cite{HM96LANL}.

To reliably extrapolate the lattice data to the continuum limit one
needs control over discretization errors and over the
matching relations between the lattice scheme and the continuum
scheme, say \MSbar.  The first issue has been addressed by the
community by simulating three different discretizations of the Dirac
action -- Wilson, SW clover, and staggered -- which have
discretization errors of $O(a)$, $O(\alpha_s(a) a)$, and $O(a^2)$
respectively. The second issue, reliability of the 
1-loop perturbative matching relations, is
being checked by using non-perturbative estimates. 

For Wilson and SW clover formulations, the internal consistency of the
lattice calculations can be checked by calculating the quark masses
two different ways. The first is based on methods of \CPT, $i.e.$ the
calculated hadron masses are expressed as functions of the quark
masses as in \CPT.  This method, based on hadron spectroscopy, is
labeled HS for brevity. In the second method, labeled WI, quark masses
are defined using the ward identity $\partial_\mu \bar \psi \gamma_5
\gamma_\mu \psi = (m_1+m_2) \bar \psi \gamma_5 \psi $.  An example of
such checks is shown in Fig.~\ref{f:Wmbar}. The solid lines are fits
to the HS and WI estimates using the Wilson action and on the same
statistical sample of configurations. The very close agreement of the
extrapolated values is probably fortuitous since the linear
extrapolation in $a$, shown in Fig.~\ref{f:Wmbar}, neglects both the
higher order discretization errors and the $O(\alpha_s(a)^2)$ errors
in the 1-loop perturbative matching relations.  The figure also shows
preliminary results for the same WI data but now with non-perturbative
Z's. The correction is large, note the large change in the slope in
$a$, yet the extrapolated value is $\lsim 4 $ \MeV. The final analysis
using non-perturbative Z's will be available soon, and it is unlikely
that the central value presented below will shift significantly.

Lastly, in the quenched approximation, estimates of quark masses can
and do depend on the hadronic states used to fix them.  The estimates
of $\mbar$ given below were extracted using the pseudoscalars mesons,
$i.e.$ pions, with the scale $a$ set by $M_\rho$.  Using either the
nucleon or the $\Delta$ to fix $\mbar$ would give $\sim 10 \%$ smaller
estimates. Similarly, extracting $m_s$ using $M_K$ gives estimates
that are $\sim 20 \%$ smaller that those using $M_{K^*}$ or $M_\phi$
as shown below. While these estimates of quenching errors are what one
would expect naively, we really have to wait for sufficient unquenched
data to quantify these more precisely.

A summary of the quenched results in \MeV at scale $\mu=2\ \GeV$,
based on an analysis of the 1997 world data \cite{Mq97rev}, is

\begin{center}
\medskip\noindent
\begin{tabular}{|l|c|c|c|}
\hline
                 &  Wilson    & TI Clover   &  Staggered    \\
\hline
$ \mbar(M_\pi)$  &  $4.1(1)$  & $3.8(1)$    & $ 3.5(1)$     \\
$ m_s(M_K)    $  &  $107(2)$  & $ 99(3)$    & $ 91(2)$      \\
$ m_s(M_\phi) $  &  $139(11)$ & $117(8)$    & $ 109(5)$  .  \\
\hline
\end{tabular}
\label{tab:mqfinal}
\end{center}
\smallskip

The difference in estimates between Wilson, tadpole improved (TI)
clover, and staggered results could be due to the neglected higher
order discretization errors and/or due to the difference between
non-perturbative and 1-loop estimates of $Z$'s. This uncertainty is
currently $\approx 15\%$.  Similarly, the $\approx 20\%$ variation in
$m_s$ with the state used to extract it, $M_K$ versus $M_{K^*}$ (or
equivalently $M_\phi$) could be due to the quenched
approximation or again an artifact of keeping only the lowest order
correction term in the extrapolations. To disentangle these 
discretization and quenching errors we again need precise unquenched data.

Thus, for our best estimate of quenched results we 
average the data and use the spread as the error. 
To these, we add a second uncertainty of $10\%$ as due to the
determination of the scale $1/a$ (another estimate of quenching errors). 
The final results, in $\MSbar$ scheme evaluated at $2 $ GeV, are~\cite{Mq97rev}
\begin{eqnarray}
\mbar &=& 3.8(4)(4)   \ \MeV  \nonumber\\
m_s   &=& 110(20)(11) \ \MeV  \,.
\label{eq:mqfinal}
\end{eqnarray}

The important question is how do these estimates change on
unquenching.  The 1996 analyses suggested that unquenching could lower
the quark masses by $\approx 20\%$~\cite{Mq96LANL,Mq96GOUGH}, however,
as discussed in \cite{Mq97rev} I no longer feel confident making an
assessment of the magnitude of the effect. The data does still
indicate that the sign of the effect is negative, $i.e.$ that
unquenching lowers the masses.  An estimate of the size requires more
unquenched data.

To end this section let me comment on a comparison of the quenched
estimates with values extracted from sum-rules as there seems to be a
general feeling that the two estimates are vastly different. In fact
the recent analyses indicate that the quenched lattice results and the
sum-rule estimates are actually consistent. A large part of the
apparent difference is due to the use of different scales at which
results are presented. Lattice QCD results are usually stated at $\mu
= 2\ \GeV$, while the sum-rules community uses $\mu = 1\ \GeV$, and the
running of the masses between these two scales is an $\approx 30\%$ effect
in full QCD. This issue is important enough that I would like to
briefly review the status of sum-rule estimates.

\section{Sum rule determinations of $\mbar$ and $m_s$}
\label{s:sumrules}

A summary of light quark masses from sum-rules is given in
Table~\ref{t:sumrules}.  Sum rule calculations proceed in one of two
ways. (i) Using axial or vector current Ward identities one writes a
relation between two 2-point correlation functions. One of these is
evaluated perturbatively after using the operator product expansion,
and the other by saturating with intermediate hadronic
states~\cite{BPR95,Jamin95}.  The quark masses are the constant of
proportionality between these two correlation functions.  (ii)
Evaluating a given correlation function both by saturating with known
hadronic states and by evaluating it
perturbatively~\cite{Narison95}. The perturbative expression depends
on quark masses, and defines the renormalization scheme in which they
are measured.  The main sources of systematic errors arise from using
(i) finite order calculation of the perturbative expressions, and (ii)
the ansatz for the hadronic spectral function. Of these the most
severe is the second as there does not exist enough experimental data
to constrain the spectral function even for $\mu < 2$ GeV. Since there
are narrow resonances in this region, one cannot match the two
expressions point by point in energy scale.  The two common approaches
are to match the moments integrated up to some sufficiently high scale
(finite energy sum rules) or to match the Borel transforms.  The hope
then is that the result is independent of this scale or of the Borel
parameter.

\begin{table} 
\begin{center}
\setlength\tabcolsep{0.25cm}
\begin{tabular}{|r|c|c|}
\hline
reference             &$\mbar$ (MeV)    &$m_s$ (MeV)        \\
\hline
\cite{Narison89} 1989 &$ {}=   6.2(0.4)$&$ {}=   138(8)    $\\
\cite{BPR95}     1995 &$ {}=   4.7(1.0)$&$                 $\\
\cite{Narison95} 1995 &$ {}=   5.1(0.7)$&$ {}=   144(21)   $\\
\cite{Jamin95}   1995 &$               $&$ {}=   137(23)   $\\
\cite{Chetyrkin} 1996 &$               $&$ {}=   148(15)   $\\
\cite{Colangelo} 1997 &$               $&$ {}=   91 - 116  $\\
\cite{Jamin97}   1997 &$               $&$ {}=   115(22)   $\\
\cite{Prades97}  1997 &$ {}=  4.9(1.9) $&$                 $\\
\cite{Yndurain}  1997 &$ {}\geq 3.8-6  $&$ {}\geq 118 - 189$\\
\cite{Dosch97}   1997 &$ {}\geq 3.4    $&$ {}\geq 88(9)    $\\
\cite{Lellouch97} 1997&$ {}\geq 4.1-4.4$&$ {}\geq 104-116  $\\
\hline
\end{tabular}
\caption{Values and bounds on $\mbar$ and $m_s$, in $\MSbar$ scheme at 
2 GeV, from sumrule analyses.}
\label{t:sumrules}
\end{center}
\end{table}


Progress in sum-rules analyses has also been incremental as in LQCD.
The perturbative expressions have now been calculated to
$O(\alpha_s^3)$~\cite{Chetyrkin}, and the value of
\(\Lambda_{QCD}^{(3)}\) has settled at $\approx 380 \MeV$.  A detailed
analysis of the convergence of the perturbation expansion suggests
that the error associated with the truncation at $O(\alpha_s^3)$ is
$\approx 10\%$ for $\mu \ge 2 $ GeV \cite{Chetyrkin}.  

Improving the spectral function has proven to be much harder.  For
example, Colangelo et al.~\cite{Colangelo} have extended the analysis
of $m_s$ in \cite{Jamin95,Chetyrkin} by constructing the hadronic
spectral function up to the first resonance ($K^*(1430)$) from known
$K \pi$ phase shift data. Similarly, Jamin~\cite{Jamin97} has used a
different parametrization of the Omnes representation of the scalar
form factor using the same phase shift data. In both cases the
reanalysis lowers the estimate of the strange quark mass
significantly. The new estimates, listed in Table~\ref{t:sumrules},
are consistent with the quenched estimates discussed in
Section~\ref{s:mfromlqcd}.


One can circumvent the uncertainties in the ansatz for the spectral
function by deriving rigorous lower bounds using just the positivity
of the spectral function \cite{SR96BGM,Yndurain,Dosch97,Lellouch97}.
Of these the most stringent are in \cite{Lellouch97} which rule out
$\mbar < 3$ and $m_s < 80$ \MeV\ for $\mu \lsim 2.5$ GeV.  The bounds,
however, have a significant dependence on the scale $\mu$ as evident by comparing the 
above values to those in the last row in Table~\ref{t:sumrules}, and the
open question is how to fix $\mu$, $i.e.$ the upper limit of
integration in the finite energy sum rules at which duality between
PQCD and hadronic physics becomes valid?  Unfortunately, this question
cannot be answered ab initio. \looseness-1

\section{Implications for $\epsilon'/\epsilon$}
\label{s_epsilon}

The Standard Model (SM) prediction 
of $\epsilon'/\epsilon$ can be written as \cite{rCP96Buras}
\begin{equation} 
\epsilon'/\epsilon = A \bigg\{c_0 + \big[c_6 B_6^{1/2} + c_8 B_8^{3/2} \big] M_r \bigg\} \ ,
\label{eq:masterEE}
\end{equation}
where $M_r = (158\MeV/(m_s + m_d))^2$ and all quantities are to be
evaluated at the scale $m_c = 1.3 \GeV$. Eq.~\ref{eq:masterEE} highlights the
dependence on the light quark masses and the bag parameters
$B_6^{1/2}$ and $B_8^{3/2}$.  For the other SM parameters that are
needed in obtaining this expression we use the central values quoted
by Buras \etal\ \cite{rCP96Buras}. Then, we get $A = 1.29\times
10^{-4}$, $c_0 = - 1.4$, $c_6 = 7.9$, $c_8 = - 4.0$. Thus, to a good
approximation $\epsilon'/\epsilon \propto M_r$. 

Conventional analysis, with $m_s + m_d = 158 \MeV$ and $B_6^{1/2} =
B_8^{3/2} = 1$, gives $\epsilon'/\epsilon \approx 3.2 \times 10^{-4}$.
The uncertainties in the remaining SM parameters used to determine $A,
c_0, c_6,$ and $c_8$ in Eq.~\ref{eq:masterEE} are large enough that,
in fact, any value between $-1 \times 10^{-4}$ and $16 \times
10^{-4}$ is acceptable\cite{rCP96Buras}.  Current experimental estimates are
$7.4(5.9)\times10^{-4}$ from Fermilab E731 \cite{epsE731} and
$23(7)\times10^{-4}$ from CERN NA31 \cite{epsNA31}.  So at present
there is no resolution of the issue whether the CKM ansatz explains
all observed CP violation.

The new generation of experiments, Fermilab E832, CERN NA48, and
DA$\Phi$NE KLOE, will reduce the uncertainty to $\approx 1
\times10^{-4}$. First results from these experiments should be
available in the next couple of years. Thus, it is very important to
tighten the theoretical prediction.

As is clear from Eq.~\ref{eq:masterEE}, both the values of quark
masses and the interplay between $B_6^{1/2}$ and $B_8^{3/2}$ will have
a significant impact on $\epsilon'/\epsilon$. The lower values of
quark masses suggested by lattice QCD analyses would increase the
estimate.  The status of results for the various B-parameters relevant
to the study of CP violation are discussed in the next section.

\section{B-parameters, $B_K$, $B_6$, $B^{3/2}_7$, $B^{3/2}_8$}
\label{s:Bparameters}

Considerable effort has been devoted by the lattice community to 
calculate the various B-parameters needed in the standard model 
expressions describing CP violation. A summary of the results 
and the existing sources of uncertainties are as follows. 

\subsection{$B_K$}
\label{s:BK}

The standard model expression for the parameter $\epsilon$, which
characterizes the strength of the mixing of CP odd and even states in
$K_L$ and $K_S$, is of the form ~\cite{CP96Buras}
\begin{equation}
|\epsilon| \sim {\rm Im}(V_{td}V_{ts}^*) B_K(\mu) \ \Phi \big({m_c\over M_W}, {m_t\over M_W}, \mu \big)\ , 
\label{eq:epsilondef}
\end{equation}
where $\Phi$ is a known function involving Inami-Lim functions and CKM elements. 
This relation provides a crucial constraint in the effort to pin down the
$\rho$ and $\eta$ parameters in the Wolfenstein parameterization of
the CKM matrix ~\cite{CP97Buras}.  The quantity $B_K$ parameterizes the QCD corrections
to the basic box diagram responsible for $K^0 - \overline{K^0}$
mixing. This transition matrix element is what we calculate on the lattice.

The calculation of $B_K$ is one of the highlights of LQCD simulations.
It was one of the first quantities for which theoretical estimates
were made, using quenched chiral perturbation theory, of the lattice
size dependence, dependence on quark masses, and on the effects of
quenching \cite{BK90PRL,BK93srs,BK96srs,BK97stag}.  Numerical data in
the staggered formulation (which has the advantage of retaining a
chiral symmetry which preserves the continuum like behavior of the
matrix elements) is consistent with these estimates in both the sign
and the magnitude. Since all these corrections have turned out to be
small, results for $B_K$ with staggered fermions have remained stable
over the last five years.

\begin{figure}[htb] 
\hbox{\hskip15bp\epsfxsize=0.7\hsize \epsfbox {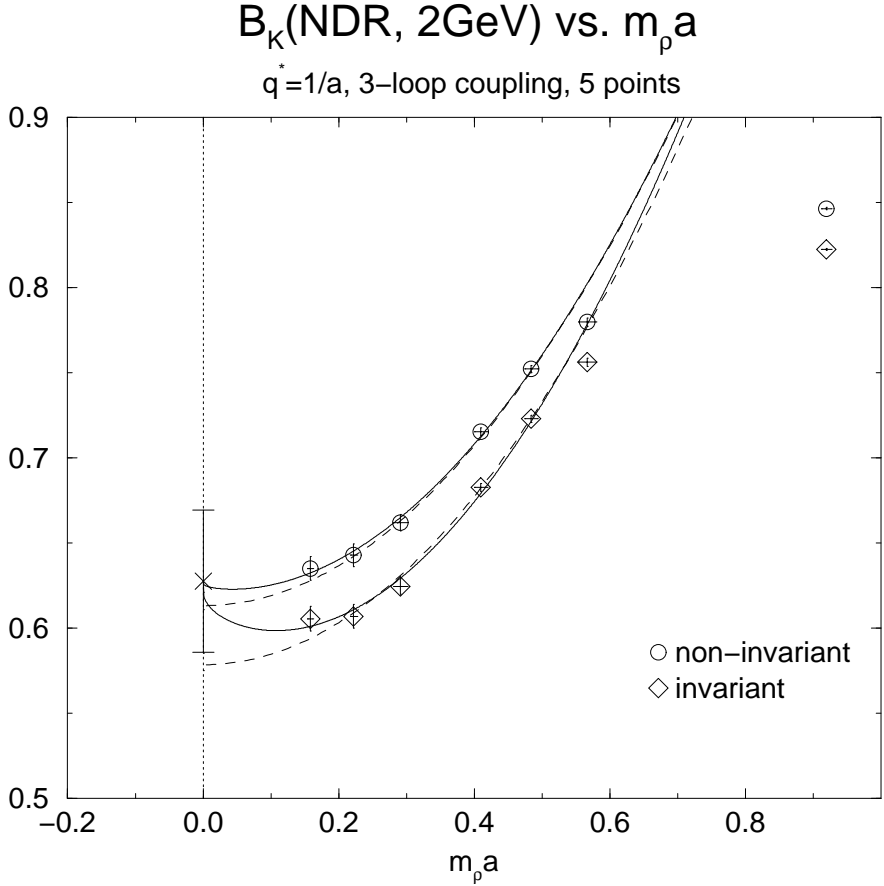}}
\caption{JLQCD data for $B_K(NDR, 2\ \GeV) $, and their extrapolation
to the continuum limit keeping both the $O(a^2)$ discretization
corrections, and the $O(\alpha_s^2)$ in the 1-loop matching factors
for two different discretizations of the weak operator. The
extrapolation of the data without including the $O(\alpha_s^2)$
corrections are shown by the dashed lines. }
\label{f-BK97JLQCD}
\end{figure}

Three collaborations have pursued calculations of $B_K$ using
staggered fermions.  The results are $B_K(NDR, 2\ \GeV) = 0.62(2)(2)$
by Kilcup, Gupta, and Sharpe \cite{BK97stag}, $0.552(7)$ by Pekurovsky
and Kilcup~\cite{B6BK97kilcup}, and $0.628(42)$ by the JLQCD
collaboration\cite{BK97JLQCD}.  Of these, the results by the JLQCD
collaboration are based on a far more extensive analysis.  The quality
of their data are precise enough to include both the leading $O(a^2)$
discretization corrections, and the $O(\alpha_s^2)$ corrections in the
1-loop matching factors.  Their data, along with the extrapolation to
$a = 0$ limit including both factors, are shown in
Fig.~\ref{f-BK97JLQCD}.  I consider theirs the current best estimate
of the quenched value.

The two remaining uncertainties in the above estimate of $B_K$ are
quenching errors and SU(3) breaking effects (all current results have
been obtained using degenerate quarks $m_u = m_d = m_s$, with kaons
composed of two quarks of mass $\sim m_s/2$ instead of $m_s$
and $m_d$).  There exists only preliminary unquenched data
\cite{BK96GK} which suggest that the effect of sea quarks is to
increase the estimate by $\approx 5\%$.  Lastly, Sharpe has used \CPT\
to estimate that the SU(3) breaking effects could also increase $B_K$
by another $4-8\%$~\cite{BK96srs}.  Confirmation of these corrections 
requires precise unquenched data which is still some years away.

In a number of phenomenological applications what one wants is the
renormalization group invariant quantity $\hat B_K$ defined, at
two-loops, as
\begin{equation}
\hat B_K = B_K(\mu)\ \big( \alpha_s(\mu) \big)^{-\gamma_0/2\beta_0} 
                     \big( 1 + {\alpha_s(\mu) \over 4 \pi} 
                       \big[ {\beta_1 \gamma_0 - \beta_0 \gamma_0 \over 2 \beta_0^2} \big] \big)\ .
\label{eq:bkhatdef}
\end{equation}
Unfortunately, to convert the quenched JLQCD number, $0.628(42)$,  one has to face the
issue of the choice of the value of $\alpha_s$ and the number of
flavors $n_f$.  It turns out that the two-loop evolution of $B_K$ is such that
one gets essentially the same number for the quenched theory,
$0.87(6)$ for $n_f=0 $ and $\alpha_s(2 GeV) = 0.192$, and $0.84(6)$
for the physical case of $n_f=3 $ and $\alpha_s(M_\tau) = 0.354$.  One
might interpret this near equality to imply that the quenching errors
are small. Such an argument is based on the assumption that there
exists a perturbative scale at which the full and quenched theories
match. Since there is no {\it a priori} reason to believe that this is
true, my preference is to double the difference and assign a second
error of $0.06$, an estimate also suggested by the preliminary
unquenched data and the \CPT\ analysis. With this caveat I arrive 
at the lattice prediction 
\begin{equation}
\hat B_K = 0.86(6)(6). 
\label{eq:bkhatans}
\end{equation}

\subsection{$\langle \pi^+\pi^0 | \CO_4 | K^+ \rangle$ and its relation to $B_K$}
\label{s:B4BK}

The $\Delta S = 2$ 
operator $\bar s L_\mu d \bar s L_\mu d $ responsible for the 
$K^0 - \overline{ K^0}$ transition belongs to the same {\bf 27} 
representation of SU(3) as the $\Delta S = 1, \Delta I = 3/2$ 
operator $\CO_4 = \bar s L_\mu d (\bar u L_\mu u - \bar d L_\mu d) + \bar s L_\mu u \bar u L_\mu d $. 
At tree level in \CPT, one gets the relation \cite{CPT97MGKL,CPT82DGH}
\begin{equation}
 \langle K^0 | \CO_{\Delta S = 2} | \overline{ K^0} \rangle 
=  {\sqrt{2} f_\pi \over 3i} \  {2 M_K^2 \over {M_K^2-M_\pi^2} }
\langle \pi^+\pi^0 | \CO_4 | K^+ \rangle 
\label{eq:KK2Kpipi}
\end{equation}
So one way to calculate $B_K$ is to measure $\langle \pi^+\pi^0 |
\CO_4 | K^+ \rangle$ on the lattice and then use
Eq.~\ref{eq:KK2Kpipi}. The motivation for doing this is that, for
Wilson-like lattice actions, $\CO_4$ is only multiplicatively
renormalized due to CPS symmetry~\cite{CPS85CBAS}, whereas the $\Delta
S = 2$ operator mixes with all other chirality operators of dimension
six. Using 1-loop values for these mixing coefficients has proven
inadequate, though the recent implementation of non-perturbative
estimates has made this situation much better~\cite{BKB7B897APE}.

The first lattice calculations of the $\Delta I = 3/2$ part of the
$K^+ \to \pi^+ \pi^0$ amplitude \cite{kpp89bernard,kpp88ape} gave
roughly twice the experimental value, even though the $B_K$ extracted
from this ``wrong'' amplitude ``agrees'' with modern lattice
estimates.  The important question therefore is how reliable are the
\CPT\ relations between
$\langle \pi^+\pi^0 | \CO_4 | K^+ \rangle\big|_{lattice}$ and 
$\langle \pi^+\pi^0 | \CO_4 | K^+ \rangle\big|_{physical}$, and between 
$\langle \pi^+\pi^0 | \CO_4 | K^+ \rangle\big|_{lattice}$ and 
$ \langle K^0 | \CO_{\Delta S = 2} | \overline{ K^0} \rangle$, 
$i.e.$ does one or both fail?

There are three sources of systematic errors in lattice calculations of $\langle
\pi^+\pi^0 | \CO_4 | K^+ \rangle$ that could explain the
contradiction. The calculation is done in the quenched approximation,
on finite size lattices, and with unphysical kinematics (in the
lattice calculation the final state pions are degenerate with the kaon
since $m_u=m_d=m_s$, and are at rest).  Of the three possibilities,
the last is the most serious as it gives a factor of two even at 
the tree-level
\begin{equation}
{2 M_K^2 \over {M_K^2-M_\pi^2}} \langle \pi^+\pi^0 | \CO_4 | K^+ \rangle \big|_{physical} = 
\langle \pi^+\pi^0 | \CO_4 | K^+ \rangle \big|_{unphysical}\ . 
\label{eq:KK2Kpipi2}
\end{equation}
This tree-level correction was taken onto account
in~\cite{kpp89bernard,kpp88ape}. The source of the remaining
discrepancy by a factor of two was anticipated by Bernard as due to
the failure of the tree-level expression~\cite{kpp89CBtasi}. Recently, Golterman and
Leung \cite{CPT97MGKL} have calculated the 1-loop corrections to
Eq.~\ref{eq:KK2Kpipi2}.  This calculation involves a number of unknown
$O(p^4)$ chiral coefficients of the weak interactions and thus has a
number of caveats. However, under reasonable assumptions about the
value of these $O(p^4)$ constants, the corrections due to finite
volume, quenching, and unphysical kinematics all go in the right
direction, and the total 1-loop correction can modify
Eq.~\ref{eq:KK2Kpipi2} by roughly a factor of two.  On the other hand
the modification to the connection with $B_K$ at the physical point is
small. 

The JLQCD Collaboration~\cite{kpp97JLQCD} has recently updated the
calculations in \cite{kpp89bernard,kpp88ape}.  By improving the
statistical errors they are able to validate the trends predicted by
1-loop \CPT\ expressions.  Thus one has a plausible resolution of the
problem. I say plausible because the calculation involves a number of
unknown chiral couplings in both the full and quenched theory and also
because of the size of the 1-loop correction. The conclusive statement
is the failure of \CPT\ for the relation Eq.~\ref{eq:KK2Kpipi2}.

The other relevant question is what bearing does the analysis of
Golterman and Leung \cite{CPT97MGKL} have on the calculations of other
$B$-parameters.  In the calculation of $B_K$, based on measuring the $
\langle K^0 | \CO_{\Delta S = 2} | \overline{ K^0} \rangle$ transition
matrix element as discussed in section~\ref{s:BK}, \CPT\ has been used
only to understand effects of finite volume and chiral logs.  These
are found to be small and the data show the predicted behavior. Based
on this success of \CPT\ we estimate that the two remaining errors --
quenching and the use of degenerate quarks -- are each $\approx 5\%$
as suggested by 1-loop \CPT. On the other hand, in present
calculations of $B_6$, $B^{3/2}_7$, and $B^{3/2}_8$ \CPT\ is used in
an essential way, $i.e.$ to relate $ \langle \pi^+\pi^0 | \CO | K^+
\rangle $ to $\langle \pi^+ | \CO | K^+ \rangle$. Second, the
calculations are done for unphysical kinematics, $i.e.$ the final
state pion is degenerate with the kaon. It would be interesting to
know the size of the one-loop corrections to these relations.

\subsection{$B^{3/2}_7$ and $B^{3/2}_8$}
\label{s:B7B8}

Assuming the reduction of $\langle \pi^+\pi^0 | \CO_{7,8}^{\Delta I =
3/2} | K^+ \rangle$ to $\langle \pi^+ | \CO_{7,8}^{\Delta I = 3/2} |
K^+ \rangle$ using tree-level \CPT\ is reliable, the lattice
calculations of $B^{3/2}_7$ and $B^{3/2}_8$ are as straightforward as
those for $B_K$.  There are three ``modern'' quenched estimates of
$B^{3/2}_7$ and $B^{3/2}_8$ which supercede all previous reported
values. These, in the NDR-\MSbar\ scheme at $2$ \GeV, are

\begin{center}
\medskip\noindent
\begin{tabular}{|l|c|c|c|c|c|}
\hline
    Fermion type                        & Matching $Z$   &  $\beta$    &  $1/a$ GeV & $B^{3/2}_7$ & $B^{3/2}_8$ \\
\hline							              
(A) Staggered \cite{BK97stag}           & 1-loop         &  $6.0,\ 6.2$&  $a \to 0$ & $0.62(3)(6)$  & $  0.77(4)(4)$ \\
(B) Wilson    \cite{B7B897wiln}         & 1-loop         &  $6.0$      &  $2.3$     & $0.58(2)(7)$  & $  0.81(3)(3)$ \\
(C) Tree-level Clover \cite{BKB7B897APE}& 1-loop         &  $6.0$      &  $2.0$     & $0.58(2)   $  & $  0.83(2)   $ \\
(D) Tree-level Clover \cite{BKB7B897APE}& Non-pert.      &  $6.0$      &  $2.0$     & $0.72(5)   $  & $  1.03(3)   $ . \\
\hline
\end{tabular}
\label{tab:B7B8}
\smallskip
\end{center}

\noindent
where I have also given the type of lattice action used, the $\beta$'s
at which the calculation was done, the lattice scale $1/a$ at which
the results were extracted, and how the 1-loop matching coefficients
were determined.

The difference between (C) and (D) is the use of perturbative versus
non-perturbative Z's.  Thus (D) is the more reliable of the APE
numbers.  The agreement between (B) and (C), in spite of the difference
in the action, is a check that the numerics are stable. All three of
these results suffer from the fact that these calculations were done
at $\beta=6.0$ ($1/a \approx 2$ GeV) and there does not yet exist data at
other $\beta$ needed to do the extrapolation to the continuum limit.

The result (A) does incorporate an extrapolation to a=0, but with only
two beta values. For example, $B^{3/2}_8= 1.24(1)$ and $1.03(2)$ at
$\beta=6.0$ and $6.2$ respectively.  Due to the large slope in $a^2$,
such an extrapolation based on two points should be considered
preliminary.  Lastly, one needs to demonstrate that corrections to
1-loop $Z's$ are under control.
 

\subsection{$B_6$}

The recent work of Pekurovsky and Kilcup \cite{B6BK97kilcup} provides the
best lattice estimate for $B_6$. Their results $B_6 = 0.67(4)(5)$ for
quenched and $0.76(3)(5)$ for two flavors have the following
systematics that are not under control. The calculation is done for
degenerate quarks, $m_u=m_d=m_s$ and uses the lowest order \CPT\ to
relate $\langle \pi\pi | \CO | K^+ \rangle $ to $\langle \pi | \CO |
K^+ \rangle $. There is no reason to believe that higher order
corrections may not be as or more significant as discussed above for $\CO_4$.
The second issue is that the 1-loop perturbative corrections in the 
matching coefficients are large. Lastly, there is no estimate for the 
discretization errors as the calculation has been done at only one value of 
$\beta$. Thus, at this point there is no solid prediction from the lattice. 

\section{Conclusions and Acknowledgements}

In view of the new generation of ongoing experiments to measure
$\epsilon'/\epsilon$ with the proposed accuracy of $1 \times 10^{-4}$,
it is very important to firm up the theoretical prediction. The
standard model estimate depends very sensitively on the sum $m_s +
m_d$ and on the interplay between the strong and electromagnetic
penguin operators, $i.e.$ $B_6^{1/2}$ and $B_8^{3/2}$. Quenched
lattice results for $(m_s + m_d)(2\ \GeV)$ are settling down at
$115(25)$ MeV, and preliminary evidence is that unquenching further
lowers these estimates.  The calculations of $B_6^{1/2}$ and
$B_8^{3/2}$ are less advanced. Hopefully we can provide reliable
quenched estimates for these parameters in the next year or
so. Thereafter, we shall start to chip away at realistic full QCD
simulations.

\bigskip
\centerline{Acknowledgements}
\bigskip

I would like to thank the organizers of CPMASS97 and ORBIS SCIENTIAE'97
for very stimulating conferences, and my collaborators Tanmoy Bhattacharya, Greg
Kilcup, and S.~Sharpe for many discussions. I also thank the DOE and
the Advanced Computing Laboratory for support of our work.

\end{document}

%% file: rgmac.latex
\newskip\defaultbaselineskip\defaultbaselineskip=12pt

\def\GeV{\mathord{\rm \;GeV}}
\def\MeV{\mathord{\rm \;MeV}}

\def\bar{\overline}

\def\eqinf{\mathrel{\raise2pt\hbox to 24pt{\raise -8pt\hbox{$t \to \infty$}\hss{$=$}}}}

\def\gsim{\mathrel{\raise2pt\hbox to 8pt{\raise -5pt\hbox{$\sim$}\hss{$>$}}}}
\def\rsim{\mathrel{\raise2pt\hbox to 8pt{\raise -5pt\hbox{$\sim$}\hss{$>$}}}}
\def\lsim{\mathrel{\raise2pt\hbox to 8pt{\raise -5pt\hbox{$\sim$}\hss{$<$}}}}
\def\ssqr#1#2{{\vbox{\hrule height.#2pt
      \hbox{\vrule width.#2pt height#1pt \kern#1pt\vrule width.#2pt}
      \hrule height.#2pt}\kern-.#2pt}}

\def\Dsl{\,\raise.15ex\hbox{$/$}\mkern-13.5mu D} 
\def\dsl{\raise.15ex\hbox{$/$}\kern-.57em\hbox{$\partial$}}

    \def\CO{{\cal O}} \def\CP{{\cal P}}

\ifx\href\undefined\def\href#1#2{{#2}}\fi
\def\spireshome{http://www.slac.stanford.edu/cgi-bin/spiface/find/hep/www?FORMAT=WWW&}
{\catcode`\%=12
\xdef\spiresjournal#1#2#3{\noexpand\protect\noexpand\href{\spireshome
                          rawcmd=find+journal+#1%2C+#2%2C+#3}}
\xdef\spireseprint#1#2{\noexpand\protect\noexpand\href{\spireshome rawcmd=find+eprint+#1%2F#2}}
\xdef\spiresreport#1{\noexpand\protect\noexpand\href{\spireshome rawcmd=find+rept+#1}}
\xdef\spireskey#1{\noexpand\protect\noexpand\href{\spireshome key=#1}}
}
\def\eprint#1#2{\spireseprint{#1}{#2}{#1/#2}}

\def\nohref{}

\def\putpaper{\edef\refpage{\the\count0}%
              \def\nohref{}%
              {\def\ {+}\def\nohref##1{}\edef\temp{\noexpand\spiresjournal
               {\journalname}{\volume}{\refpage}}\expandafter}\temp
               {\sfcode`\.=1000{\journalname} {\bf \volume} (\refyear)
                \refpage}\egroup}
\def\putpage{\edef\refpage{\the\count0}%
              \def\nohref{}%
              {\def\ {+}\def\nohref##1{}\edef\temp{\noexpand\spiresjournal
               {\journalname}{\volume}{\refpage}}\expandafter}\temp
              {\refpage}\egroup}
\def\dojournal#1#2 (#3){\def\journalname{#1}\def\volume{#2}\def\refyear
                        {#3}\afterassignment\putpaper\bgroup\count0=}
\def\morepage{\afterassignment\putpage\bgroup\count0=}

\def\NPB#1{\dojournal{Nucl.\ Phys.}{B#1}}
\def\NPBPS#1{\dojournal{Nucl.\ Phys.\ \nohref(Proc.\ Suppl.\nohref)}{\nohref B#1}}

\def\PRL#1{\dojournal{Phys.\ Rev.\ Lett.}{#1}}
\def\PR#1{\dojournal{Phys.\ Rev.}{#1}}
\def\PRep#1{\dojournal{Phys.\ Rep.}{#1}}

\def\PRD#1{\dojournal{Phys.\ Rev.}{D#1}}

\def\PLB#1{\dojournal{Phys.\ Lett.}{#1B}}
\def\RMP#1{\dojournal{Rev.\ Mod.\ Phys.}{#1}}

\def\ZPC#1{\dojournal{Z.\ Phys.}{C#1}}

\def\etal{{\it et al.}}

